\documentclass[a4paper]{article}
\usepackage[latin9]{inputenc}
\usepackage{booktabs}
\usepackage{multirow}
\usepackage{amsmath}
\usepackage{graphicx}
\usepackage{rotating}
\makeatletter

\usepackage{balance,setspace}
\usepackage{multirow,comment, color}
\usepackage{amsmath}
\usepackage{soul}
\usepackage{arydshln}
\usepackage{enumitem}
\newlength\myindent
\def\Width{0\kern2\tabcolsep\ldots\kern1\tabcolsep0}
\usepackage{etoolbox}
\newcommand{\zerodisplayskips}{%
  \setlength{\abovedisplayskip}{3pt}
  \setlength{\belowdisplayskip}{3pt}
  \setlength{\abovedisplayshortskip}{3pt}
  \setlength{\belowdisplayshortskip}{3pt}}
\appto{\normalsize}{\zerodisplayskips}
\appto{\small}{\zerodisplayskips}
\appto{\footnotesize}{\zerodisplayskips}

\providecommand{\tabularnewline}{\\}

\usepackage{INTERSPEECH2020}

\pdfoutput=1

\title{Complementary Language Model and Parallel Bi-LRNN for False Trigger Mitigation}
\name{Rishika Agarwal, Xiaochuan Niu, Pranay Dighe, Srikanth Vishnubhotla, Sameer Badaskar, Devang Naik}
\address{Apple, One Apple Park Way, Cupertino, CA, USA}
\email{\{rishika\_agarwal,xniu,pdighe,svishnubhotla,badaskar,naik.d\}@apple.com}

\makeatother

\begin{document}

\maketitle %
\begin{abstract}
\vspace{-1mm}
False triggers in voice assistants are unintended invocations
of the assistant, which not only degrade the user experience but may
also compromise privacy. False trigger mitigation (FTM) is a process to detect the false trigger events and respond appropriately
to the user. In this paper, we propose a novel solution to the FTM
problem by introducing a parallel ASR decoding process with a special
language model trained from ``out-of-domain'' data sources. Such language model is complementary to the existing language model
optimized for the assistant task. A bidirectional lattice RNN (Bi-LRNN)
classifier trained from the lattices generated by the complementary
language model shows a $38.34\%$ relative reduction of the false trigger
(FT) rate at the fixed rate of $0.4\%$ false suppression (FS) of
correct invocations, compared to the current Bi-LRNN model. In addition, we propose to train a parallel Bi-LRNN model based on the decoding lattices from both language models, and examine various ways of implementation. The resulting model leads to further reduction in the false trigger rate by $\ensuremath{10.8\%}$. 
\end{abstract}


{\let\thefootnote\relax\footnote{{To appear in Proceedings of InterSpeech 2020}}}

\vspace{-2mm}
\section{Introduction}
\vspace{-1mm}
Voice trigger detection is a vital part of current voice assistant
products. In such systems, one or multiple trigger phrases are defined
for users to invoke the device to process voice requests. The design
of a trigger detector is often constrained by limited computation
resources and power consumption of hardware, therefore we often adopt
simple DSP and acoustic models\,\cite{SigtiaEtal:2018:IS,Wu:2018:ICASSP}. In practice, a trigger detector 
is usually operated in a low-false-rejection mode in order to allow
most acoustic samples to be passed to downstream processes. However,
such design may cause
the assistant (wrongly) respond to unintended acoustic inputs. There
are also cases when users accidentally invoke the assistant through
a UI element such as a button-press or a particular gesture. Such
unintended invocations of voice assistants can be referred
to as ``false triggers''. To mitigate the false trigger cases, one
can introduce an extra process to determine whether an
acoustic sample is intended or not, which is in essence a binary classification
problem.

The false trigger mitigation process can make use of both acoustic
and linguistic clues from the input sample. When the errors are due
to voice trigger detector, an intuitive approach is to feed
the acoustic sample into an ASR system and check for the existence
of trigger phrases in the 1-best output\,\cite{MichaelyEtal:2017:ASRU}. In more
general cases, the text output contains the intent information
from the user, therefore can be used as input to the classifier.
In\,\cite{MallidiEtal:2018:IS}, the 1-best output is encoded as
an LSTM embedding to represent the linguistic feature. It is combined
with the LSTM embedding of the acoustic features, and decoder features
including trellis entropy, Viterbi cost, confidence and average number
of arcs as the final input feature set to the classier. Considering
the ASR results may contain errors, the decoder features are designed
explicitly to capture the ambiguity during the decoding process. A
recent follow-up work\,\cite{HuangEtal:2019:IS} focuses on improving
the acoustic features by incorporating utterance-level representations.
It also introduces dialog-type information to facilitate the classifier to make better decisions.

To build an intent classifier, the authors of\,\cite{LadhakEtal:2016:IS} propose a condense representation
of lattices from ASR decoder, called ``Lattice RNN'' (LRNN). By
introducing a pooling operation over the incoming arcs of each node in the lattice,
and a propagation operation over the outgoing arcs of the nodes, the
authors are able to construct a neural network on a lattice, and encode
the whole lattice information as the vector output from the final
node of the lattice. The LRNN embedding is used as the input vector of the intent
classifier, which achieves better accuracy and faster run-time, compared to the baseline model running on N-best results. A similar approach
can be applied to the FTM task. Our previous work\,\cite{JeonLiuMason:2019:ICASSP} redefined the feature set attached to each arc in the decoder lattice, and extended the network to bi-directional (Bi-LRNN). The decoding lattice is encoded as the concatenation of hidden layers from the start and end nodes of the lattice. Thus a classifier built on top of the Bi-LRNN is able to mitigate the false trigger cases significantly.
A recent work\,\cite{PranayEtal:2020:ICASSP} explored the use of
graph neural networks (GNN) to encode the decoding lattice, which
achieves similar accuracy as the Bi-LRNN representation with more
efficient training.

In this paper, we investigate the impact of the decoder's language
model (LM) on false trigger mitigation. Considering that
the voice assistant's LM is usually well trained with \emph{in-domain}
data, and the LM also tends to see more usage data with the trigger
phrase at the beginning, it is likely that the LM is biased towards
the \emph{in-domain} data, which thereby biases it towards detecting
the trigger phase. This bias may reduce the power of the decoding
lattice in mitigating false triggers. In our study, we train a new
LM that is not biased to the trigger phrase and \emph{in-domain} data.
We compare the mitigation performance between the Bi-LRNN classifiers
built from the the lattice outputs of different LMs. We further
investigate how to make use of the complementary information in two
different language models, and propose some approaches to build parallel
Bi-LRNN, which leads to further improvement in false trigger mitigation.

\vspace{-2mm}
\section{Method}
\vspace{-1mm}
\subsection{Bi-LRNN for false trigger mitigation}
\vspace{-1mm}

In our baseline Bi-LRNN system, we obtain the word hypothesis lattice
$L$ for an acoustic sample $X$, from the ASR decoder. The lattice consists of a start node,
an end node, and other intermediate nodes. The nodes are connected via arcs and each arc has a feature vector associated with it. 
The Bi-LRNN computes a forward and backward latent embedding for each node in the lattice (refer to \cite{JeonLiuMason:2019:ICASSP} for more details).

The final outputs of the Bi-LRNN are the forward latent embedding of the end node $\textbf{h}_{f}(s_{end})$ and backward latent embedding of the start node $\textbf{h}_{b}(s_{start})$, where $s_{start}$ and $s_{end}$ denote the lattice's start and end nodes. A feed-forward classifier then takes the input as $[\textbf{h}_{f}(s_{start}),\textbf{h}_{b}(s_{end})]$.
The classifier gives a real valued output $y\in[0,1]$, which is converted
to a label $l_{pred}\in\{0,1\}$ by choosing a threshold $t$. The
threshold can be kept fixed at certain value, or can be evaluated
empirically on the cross validation set, to achieve the desirable
False Suppression (FS) of invocation rate.

\vspace{-2mm}
\subsection{Parallel decoding with complementary LMs}
\vspace{-1mm}
A typical ASR decoding process can be formulated as searching the
best word sequence $W^{*}$ that maximizes \eqref{eq:asr-1},
where $P\left(X\mid W\right)$ denotes acoustic model (AM), representing
the conditional probability of acoustic features $X$ given a word
sequence $W$, and $P\left(W\right)$ denotes language model (LM),
representing the probability of any word sequence $W$. Ideally, the
LM of an ASR system should approximate the distribution of all the
word sequences that could reach the decoder. However, in practice,
the voice assistant application is only designed to respond to a relevant
set of user requests. So the LM is usually trained to maximize the
likelihood of \emph{in-domain} sentences. If we refer to the \emph{in-domain}
sentences as a class $\mathcal{L}_{D}$, and \emph{out-of-domain}
sentences as a class $\mathcal{L}_{O}$, the ASR LM trained from \emph{in-domain}
data can be explicitly represented as $P\left(W\mid\mathcal{L}_{D}\right)$
in \eqref{eq:asr-2}, with $P\left(\mathcal{L}_{D}\right)$
denoting the prior probability of in-domain usage.
\begin{align}\small
W^{*} & =\arg\underset{W}{\max}\left\{ P\left(X\mid W\right)P\left(W\right)\right\} \label{eq:asr-1}\\
 & =\arg\underset{W}{\max}\left\{ P\left(X\mid W,\mathcal{L}_{D}\right)P\left(W\mid\mathcal{L}_{D}\right)P\left(\mathcal{L}_{D}\right)\right\}\label{eq:asr-2}
\end{align}
To use the ASR decoding information to determine whether a sequence
of acoustic features represent an unintended invocation, we can compute
the probability of \emph{in-domain} usage given the acoustic observation,
$P\left(\mathcal{L}_{D}\mid X\right)$. This measurement can be expanded
as in \eqref{eq:asr-3}, in which the first factor is the
summation of AM and LM probabilities over all sentence hypotheses. An
approximation can be made to apply the summation over the resulting
lattice paths during decoding, when ignoring the low likelihood word
sequences being pruned. The Bi-LRNN embedding can be interpreted as
an implicit representation of such measurement with more flexibility
and modeling capacity.\,\cite{JeonLiuMason:2019:ICASSP}
\begin{align}\small 
P\left(\mathcal{L}_{D}\mid X\right)&=P\left(X\mid\mathcal{L}_{D}\right)P\left(\mathcal{L}_{D}\right)/P\left(X\right)\\
=&\frac{\sum_{i}\left\{ P\left(X\mid W_{i},\mathcal{L}_{D}\right)P\left(W_{i}\mid\mathcal{L}_{D}\right)\right\} P\left(\mathcal{L}_{D}\right)}{P\left(X\right)}\label{eq:asr-3}
\end{align}

The drawback of the above measurement in \eqref{eq:asr-3}
is that $P\left(W\mid\mathcal{L}_{D}\right)$ only contains \emph{in-domain}
information, so its power of rejecting false triggered samples may
be limited. If we have a good estimation of \emph{out-of-domain} sentences
with LM $P\left(W\mid\mathcal{L}_{O}\right)$, we can construct a
complementary measurement, $P\left(\mathcal{L}_{O}\mid X\right)$,
which in theory should have more power to reject false trigger. Equation\,\eqref{eq:asr-4}
implies we will run ASR decoding with a different set of LM, $P\left(W\mid\mathcal{L}_{O}\right)$,
to generate lattices different from the default ones. We can apply
the similar Bi-LRNN operation on the \emph{out-of-domain} lattices
for more modeling capacity.
\begin{equation}\small \label{eq:asr-4}
P\left(\mathcal{L}_{O}\mid X\right)=\frac{\sum_{i}\left\{ P\left(X\mid W_{i},\mathcal{L}_{O}\right)P\left(W_{i}\mid\mathcal{L}_{O}\right)\right\} P\left(\mathcal{L}_{O}\right)}{P\left(X\right)}
\end{equation}
Furthermore, we can derive a probability ratio measurement as shown
in \eqref{eq:ratio}, which adopts the ratio between the
\emph{in-domain} and \emph{out-of-domain} probabilities given the
acoustic observation to balance the suppression/trigger decision.
This measurement implies two ASR decoders can be run in parallel to
achieve two different lattices from the same acoustic input. By combining
the information from both lattices, we may be able to achieve better
discriminative capacity between the two classes. Once more, this measurement
can be generalized by training two Bi-LRNNs from the lattices of two
decoders, and hopefully the network can learn more complex relationship
between the two lattices when the targeting cost function is set to
minimize the classification errors.
\begin{equation}\small \label{eq:ratio}
\frac{P\left(\mathcal{L}_{D}\mid X\right)}{P\left(\mathcal{L}_{O}\mid X\right)}=\frac{\sum_{i}\left\{ P\left(X\mid W_{i},\mathcal{L}_{D}\right)P\left(W_{i}\mid\mathcal{L}_{D}\right)\right\} P\left(\mathcal{L}_{D}\right)}{\sum_{j}\left\{ P\left(X\mid W_{j},\mathcal{L}_{O}\right)P\left(W_{j}\mid\mathcal{L}_{O}\right)\right\} P\left(\mathcal{L}_{O}\right)}
\end{equation}

\vspace{-2mm}
\subsection{Ensembling parallel Bi-LRNNs\label{subsec:Ensembling-parallel-LRNNs}}
\begin{figure}[t]
\centering \includegraphics[width=0.72\linewidth]{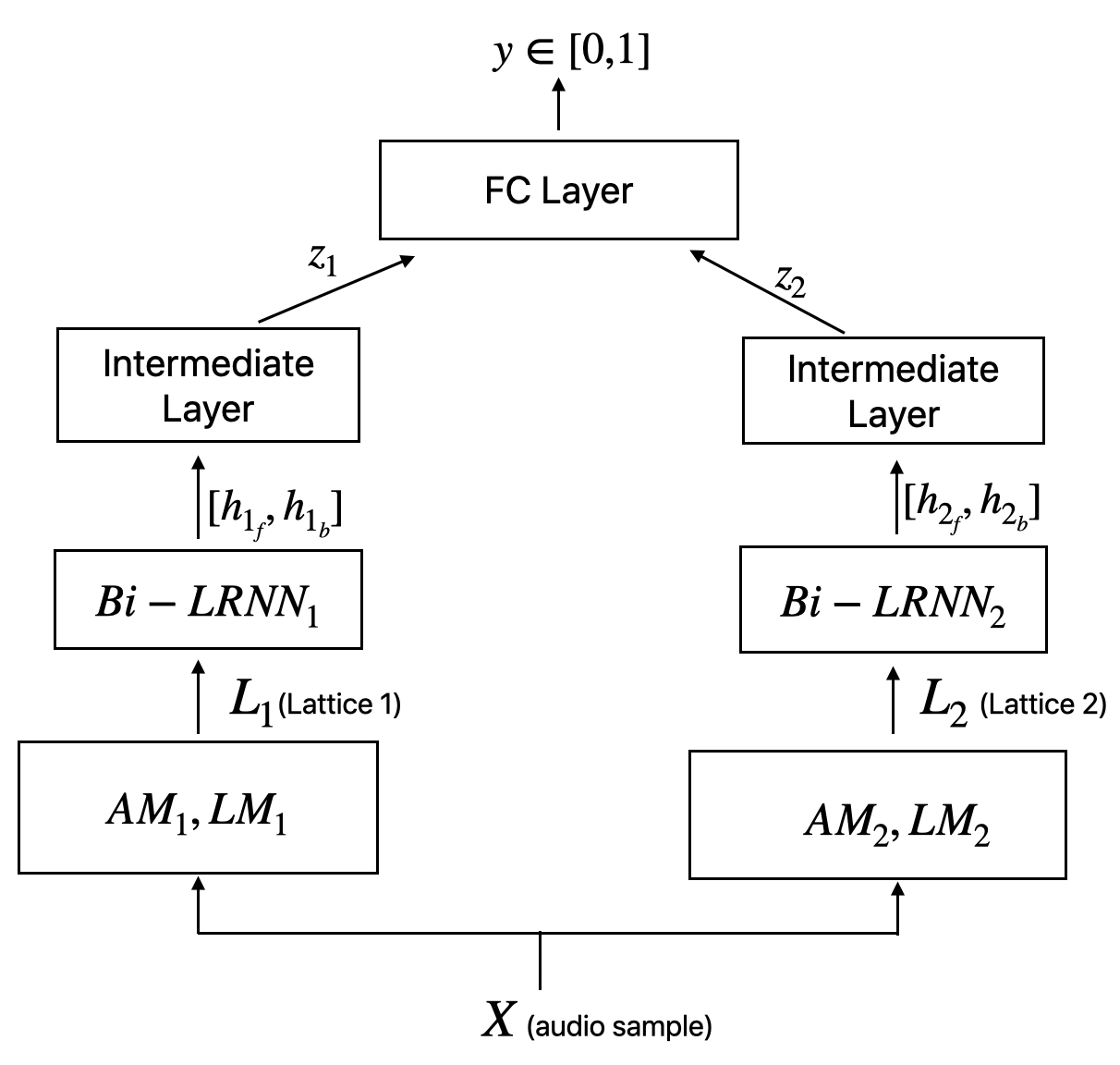}
\caption{Schematic diagram of Parallel LRNN model}
\label{fig:backtrain} 
\end{figure}

\begin{figure}[t]
\includegraphics[width=1\linewidth]{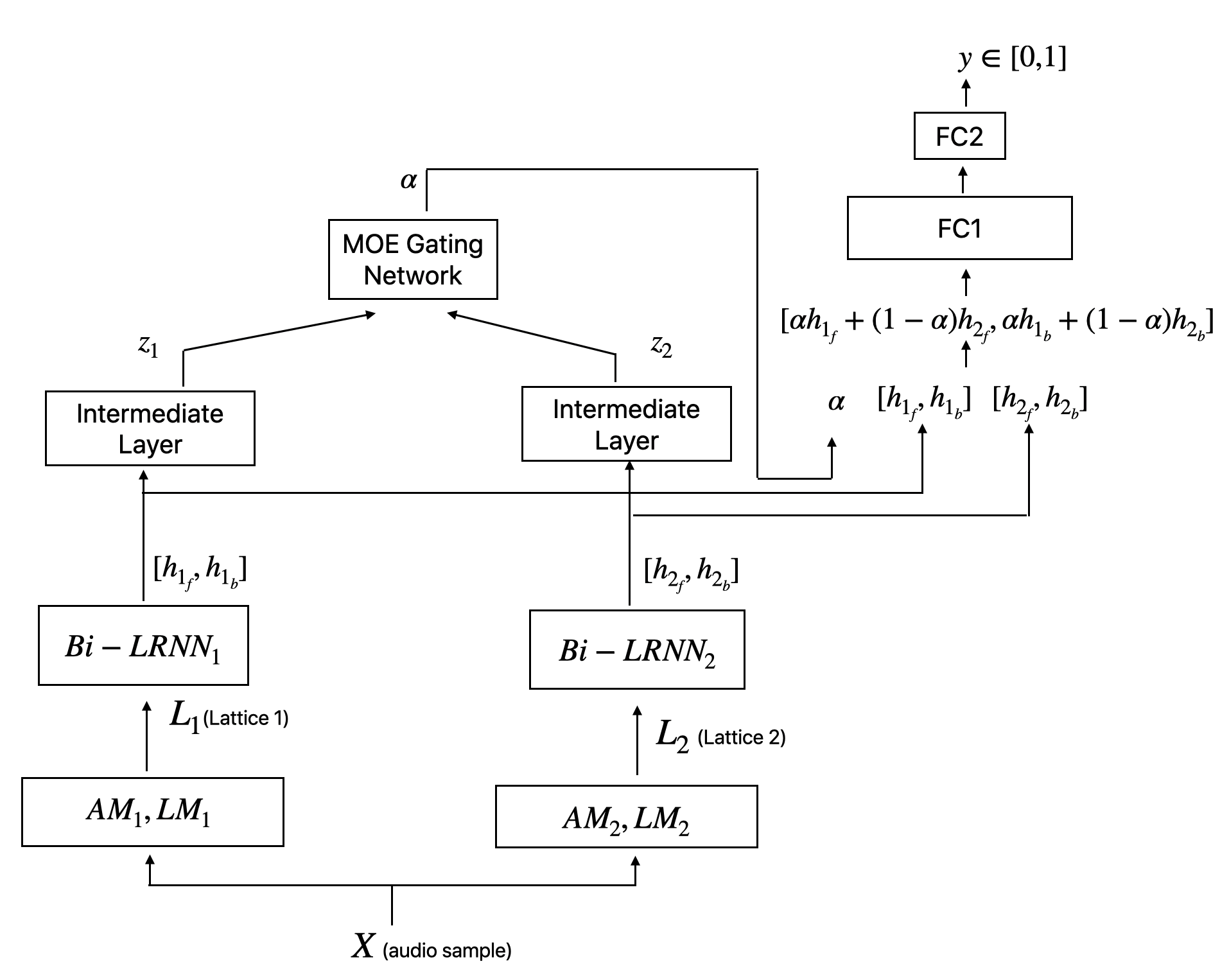} \caption{Schematic diagram of Mixture of Experts model}
\label{fig:moe} 
\end{figure}

\vspace{-1mm}
In the error analysis (Section 3.4), we show that the base model is
more accurate for some examples, and the out-of-domain model is better
on others, depending on the true label of the example. Thus, the language
models likely represent complementary information, and a model comprised
of both the LMs could out-perform the individual models based on either
of the LMs. To achieve this, the outputs from the two Bi-LRNN models
can be combined in different ways before passing to the classifier.
We explore the following ensembling techniques, and compare the FT rates
achieved by each of them in Section 3.5:
\begin{itemize}[topsep=0pt,itemsep=0pt]
\item \textbf{Combine scores from the pre-trained Bi-LRNNs}: We take the prediction
scores $y_{1}$ and $y_{2}$ from the two Bi-LRNNs trained separately,
and pass them to a shallow classifier. (Only classifier layers are
trained). 
\item \textbf{Combine the Bi-LRNN embeddings from the pre-trained Bi-LRNNs}: We take
the latent Bi-LRNN embeddings $\textbf{h}_{1_{f}},\textbf{h}_{1_{b}},\textbf{h}_{2_{f}},\textbf{h}_{2_{b}}$
from the pre-trained Bi-LRNNs, and pass them to a classifier (Here
again, only the classifier is trained). 
\item \textbf{Train the Bi-LRNNs in parallel, by back-propagating the classifier
loss}: The setting is the same as the previous case, but we back-propagate
the classifier loss to both the Bi-LRNNs as well. Thus, the entire
model is trained end-to-end (from scratch or by loading the weights
of the trained Bi-LRNNs and fine-tuning them). The schematic of the
model is shown in Figure\,\ref{fig:backtrain}. 
\item \textbf{Mixture of Experts}: Instead of concatenating the embeddings of the
two Bi-LRNNs, we can pass their weighted sum to the classifier. A
Mixture of Experts model\,\cite{JabobsEtal:NeruComp:1991}, computes
the relative importance of each ``expert'' (in this case, the two
Bi-LRNNs are the ``experts''), and weighs the outputs of the models
by a parameter $\alpha$. The weight parameter $\alpha$ determines
the reliability of each Bi-LRNN for an input sample, and we pass a
weighted sum of the lattice embeddings to a classifier $\left(\left[\alpha\mathbf{h}_{1_{f}}+\left(1-\alpha\right)\mathbf{h}_{2_{f}},\alpha\mathbf{h}_{1_{b}}+\left(1-\alpha\right)\mathbf{h}_{2_{b}}\right]\right)$.
The model is trained end-to-end. The schematic of the model is shown
in Figure\,\ref{fig:moe}. 
\end{itemize}
\vspace{-2mm}
\section{Experiment and results}
\vspace{-1mm}
\subsection{FTM Dataset and evaluation metrics}
\vspace{-1mm}
All our experiments are performed on an FTM dataset\,\cite{PranayEtal:2020:ICASSP},
which is composed of far field usage samples with manual labels of
``true trigger'' (TT) and ``false trigger'' (FT) classes. The
raw audio data are split into \emph{train}, \emph{cv}, \emph{dev},
and \emph{eval} sets for the purposes of training, cross-validation,
development and evaluation. The \emph{train} and \emph{cv} sets are
augmented by adding gain, noise, and speed perturbations, which increases
the amount of training data by 3x. Table\,\ref{tab:dataset} summarizes
the amount of data in each set and condition.

We train FTM classifiers for multiple epochs on the \emph{train} set.
The training epoch which achieved the lowest FT on the \emph{cv} set
is evaluated on the \emph{dev} and \emph{eval} sets. 
We expect our voice assistant to have minimal false triggers, and maximum true positives (minimal FS), for a good user-experience. We thus focus on the low FS regime in our DET curves, and the lower the AUC (Area Under Curve) of the DET curve, the better the model. 
In our experiments, we arbitrarily choose a small FS rate of $0.4\%$ to act as the operating point. So the
False Trigger (FT) rate at this FS rate is the key metric in evaluating
the false trigger mitigation models, while the AUC gives us an estimate of how good the model performs overall, irrespective of the operating point. We set the threshold that achieves the target FS rate (0.4\%) on the \emph{dev} set. The performance metric of concern to us is the corresponding FT rate on
the \emph{eval} set. 

\begin{table}[b]
\centering %
\begin{tabular}{c|cccc}
\hline 
\textbf{Label}  & \textbf{train}  & \textbf{cv}  & \textbf{dev}  & \textbf{eval}\tabularnewline
\hline 
\textbf{True Trigger}  & $14,225\times3$  & $1,582\times3$  & $5,829$  & $11,646$\tabularnewline
\textbf{False Trigger}  & $6,223\times3$  & $691\times3$  & $5,657$  & $11,316$\tabularnewline
\hline 
\end{tabular}
\caption{\label{tab:dataset} \footnotesize FTM dataset for model training and evaluation}
\end{table}
\vspace{-2mm}
\subsection{ASR decoder and baseline models}
\vspace{-1mm}

In all experiments, we adopt an internal ASR decoder with various
model configurations. The acoustic model has an Hidden Markov model
(HMM) and Convolutional Neural Network (CNN) hybrid structure\,\cite{Huangetal:ICASSP},
which is trained with filter bank features from US English speech
data using cross-entropy and subsequent BMMI objective functions\,\cite{VeselyEtal:2013:IS}.
The CNN comprises 50 layers and uses the scaled exponential linear
unit (SELU) activation function to achieve self-normalization during
training\,\cite{KlambauerEtal:2017:NIPS}, which achieves state-of-art
performance. The baseline language model in the decoder is a 4-gram
model interpolated from multiple sub-LMs trained from different data
sources that are relevant to the far field application (in-domain).
The data sources include enumerations of various usage domains, the
re-decoding transcripts of live usage, and accumulated error corrections
from the users. All the sub-LMs share a word lexicon with a vocabulary
size of around $570K$. The final interpolated LM is pruned to contain
about $4.9M$ 4-grams, $8.0M$ trigrams and $4.8M$ bigrams. We refer
to this as the \emph{BaseLM}.

\vspace{-2mm}
\subsection{Bi-LRNN based on complementary LM}
\vspace{-1mm}
In order to capture the out-of-domain usages, we consider the following
data sources to train a complementary language model called \emph{ChatterLM}.
The first data source is from the automatic transcriptions of the
dictation application; the second source is from the voice search
application. The language usage styles of these two applications are
different from that of the assistant application in our current study.
The third source is artificial data generated from enumeration of extra use cases that are not relevant to the specific device under study. The \emph{ChatterLM} is built in the same way as the \emph{BaseLM} in production, then combined
with the baseline AM for the ASR decoder to use.

\begin{figure}
\centering\includegraphics[width=0.9\linewidth]{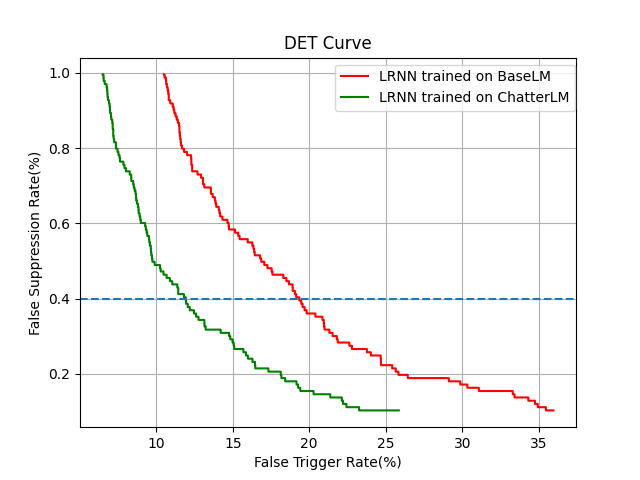}
\caption{\footnotesize DET curves of \emph{BaseLM} and \emph{ChatterLM} LRNN models.}
\label{fig:roc-chatter} 
\end{figure}

With the two sets of ASR models, we generate decoding lattices on
the FTM \emph{train} and \emph{cv} sets, then build two separate Bi-LRNN
classifiers from the lattice features. To compare the accuracy between
the two classifiers, we plot the DET curves of them on the \emph{eval}
sets in Figure\,\ref{fig:roc-chatter} (We plot only the region of interest of the DET curve, ie $FS < 1\%$). At the operation points around the fixed FS rate of $0.4\%$, the \emph{ChatterLM} based classifier
achieves a lower FT rate than the baseline classifier based on \emph{BaseLM}. The relative
reduction of FT rate is $38.34\%$ (FT reduces from $19.35\%$ for
the \emph{BaseLM} based Bi-LRNN to $11.93\%$ for the \emph{ChatterLM} based Bi-LRNN).
Such a significant FT rate reduction clearly indicates that the LM
trained from out-of-domain data sources is more capable of detecting
false triggers than the LMs trained from in-domain data sources.

\vspace{-2mm}
\subsection{Error Analysis}
\vspace{-1mm}
We choose the two Bi-LRNN classifiers, which use lattices from
the \emph{BaseLM} and the \emph{ChatterLM} respectively, to further
analyze the error patterns. The idea is that if the models make mistakes
on different samples, then ensembling them would provide complementary
information, and thus improve the overall performance. We compute
the matrix showing the number of samples for which the \emph{BaseLM}
Bi-LRNN and the \emph{ChatterLM} Bi-LRNN got the predictions correct
and incorrect (see Table\,\ref{tab:FTM_error}). Both models achieve
very high accuracy on the True Trigger (TT) class, with \emph{BaseLM}
being the more accurate of the two. For the False Trigger (FT) class,
both models are less accurate, and \emph{ChatterLM} is more accurate
than \emph{BaseLM} (only $3.66\%$ samples where \emph{BaseLM} Bi-LRNN
gets correct and \emph{ChatterLM} Bi-LRNN gets wrong, cf. $20.27\%$
samples where \emph{BaseLM} Bi-LRNN gets wrong and \emph{ChatterLM}
Bi-LRNN gets correct). These results align with our expectations,
as the \emph{ChatterLM} Bi-LRNN model is expected to be more accurate
for unintended speech samples since it uses an LM trained on out-of-domain
data, while the \emph{BaseLM} Bi-LRNN uses the LM primarily trained
on in-domain data. Thus, the models are stronger in different sample
spaces, and should be able to complement each other when used together
in an ensemble model.

\begin{table}[b]\footnotesize
\centering %
\begin{tabular}{cccc}
\toprule 
\begin{turn}{90}
\end{turn} &  & \textbf{ChatterLM Correct}  & \textbf{ChatterLM Wrong}\tabularnewline
\midrule 
\multirow{2}{*}{\textbf{TT}} & \textbf{BaseLM Correct}  & $99.4\%$  & $0.39\%$\tabularnewline
 & \textbf{BaseLM Wrong}  & $0.08\%$  & $0.09\%$\tabularnewline
\midrule 
\multirow{2}{*}{\textbf{FT}} & \textbf{BaseLM Correct}  & $66.3\%$  & $3.66\%$\tabularnewline
 & \textbf{BaseLM Wrong}  & $20.27\%$  & $9.76\%$\tabularnewline
\cmidrule{2-4} \cmidrule{3-4} \cmidrule{4-4} 
\end{tabular}
\caption{\label{tab:FTM_error}\footnotesize Error Analysis of True (TT) and False (FT) Triggers}
\end{table}
\vspace{-2mm}
\subsection{Parallel Bi-LRNNs}
\begin{figure}
\centering\includegraphics[width=1.1\linewidth]{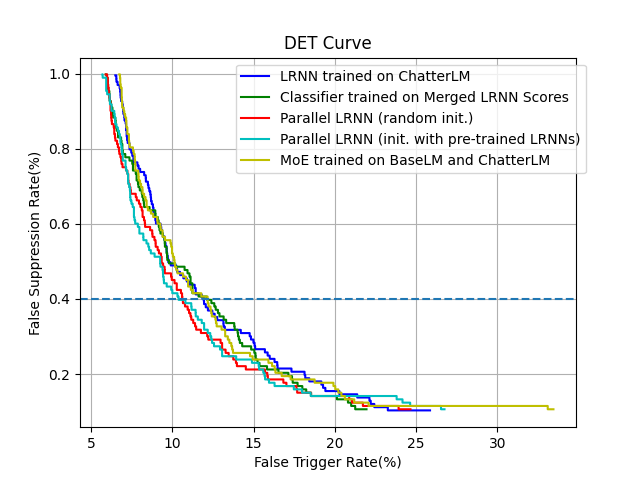}
\caption{\label{fig:ROC-curves-of-ParrallLRNN} \footnotesize DET curves of \emph{ChatterLM} and Ensemble of parallel Bi-LRNNs trained on \emph{BaseLM} and \emph{ChatterLM}}
\end{figure}

\vspace{-1mm}
Assuming we have two ASR models available, one comprising of the LM
trained on in-domain data (\emph{BaseLM}), and the other comprising
of LM trained on out-of-domain data (\emph{ChatterLM}), we can leverage
the two complementary lattices by training parallel Bi-LRNN classifiers.
We implement the different ensembling methods proposed in Section\,\ref{subsec:Ensembling-parallel-LRNNs}
to compare their performance.  Figure\,\ref{fig:ROC-curves-of-ParrallLRNN} shows
the DET curves (restricted to the region of interest) of different parallel Bi-LRNN models, along with that of the single \emph{ChatterLM} based Bi-LRNN. Table\,\ref{tab:results} shows the
FT rates of these classifiers at the fixed FS rate of $0.4\%$, and the Area under the DET curve.
\begin{table}[th]\footnotesize
\centering %
\begin{tabular}{lll}
\toprule 
\textbf{Classifier}  & \textbf{FT at FS = 0.4\%} & \textbf{AUC} \tabularnewline 
\midrule 
\emph{BaseLM} based Bi-LRNN  & $19.35\%$ & $0.0067$ \tabularnewline
\emph{ChatterLM} based Bi-LRNN  & $11.93\%$  & $0.0041$ \tabularnewline
\midrule 
Classifier on merged scores  & $12.4\%$  & $\textbf{0.0037}$ \tabularnewline
Classifier on merged embedding vectors  & $13.83\%$  & $0.0087$\tabularnewline
Fully trained parallel Bi-LRNN  & \tabularnewline
(random initialization)  & $\mathbf{10.64}\%$  & $0.0049$ \tabularnewline
Fully trained parallel Bi-LRNN  &  \tabularnewline  
(initialized with pre-trained weights)  & $10.80\%$  & $0.0043$ \tabularnewline
Mixture of Experts  & $12.18\%$  & $0.0051$ \tabularnewline
\bottomrule
\end{tabular}
\caption{\label{tab:results} \footnotesize False Trigger rates for different models}
\end{table}
Fully trained parallel Bi-LRNNs achieve a better FT rate than the \emph{ChatterLM} based single Bi-LRNN classifier,
while the classifiers trained on merged scores or embeddings, and the Mixture of Experts model perform better than the BaseLM based Bi-LRNN classifier, but worse than the ChatterLM based classifier alone. The best performance is achieved by the classifier trained by fully back-propagating the loss to the Parallel
Bi-LRNNs -- $\ensuremath{10.8\%}$ relative reduction in FT rate (over the \emph{ChatterLM} based
Bi-LRNN baseline). Initializing the Bi-LRNNs with individually pre-trained
Bi-LRNNs gives almost identical results as random initialization (red and cyan curves in Fig \ref{fig:ROC-curves-of-ParrallLRNN}); At the operating point (FS = 0.4\%), fine-tuning the pre-trained Bi-LRNNs is slightly worse than training from random initialization, although the former has marginally lower AUC. The improvement made by parallel Bi-LRNN model over
the single \emph{ChatterLM} based Bi-LRNN is consistently significant in our region of interest, ie,
for FS rates below $\ensuremath{1\%}$.

\vspace{-2mm}
\section{Conclusions}
\vspace{-1mm}
We proposed a novel solution to the ASR lattice based
false trigger mitigation approach by introducing a complementary LM
to the decoding process. The LM is trained from out-of-domain
data sources and provides complementary information to
the original LM optimized for in-domain ASR accuracy. We demonstrated
that a Bi-LRNN classifier built from the lattices generated from the
complementary LM significantly outperforms the classifier
built from the baseline ASR model set. With this single \emph{ChatterLM}
Bi-LRNN, we achieved a $38.34\%$ relative reduction of the FT rate
at the fixed $0.4\%$ FS level comparing to the current production
FTM model. Furthermore, we proposed a novel approach of parallel Bi-LRNN,
and examined multiple ways to implement and train the classifier.
By back-propagating the training loss fully
to the parallel Bi-LRNN network, we saw a further $\ensuremath{10.8\%}$ relative reduction
of the FT rate. These
results indicate that there is room for improving the traditional
ASR decoder in the FTM task, and encourage us to reconsider the architecture
design that can enable parallel LM decoding and parallel Bi-LRNN computation.

\bibliographystyle{IEEEtran}
\bibliography{refs}

\begin{thebibliography}{10}
\providecommand{\url}[1]{#1}
\csname url@samestyle\endcsname
\providecommand{\newblock}{\relax}
\providecommand{\bibinfo}[2]{#2}
\providecommand{\BIBentrySTDinterwordspacing}{\spaceskip=0pt\relax}
\providecommand{\BIBentryALTinterwordstretchfactor}{4}
\providecommand{\BIBentryALTinterwordspacing}{\spaceskip=\fontdimen2\font plus
\BIBentryALTinterwordstretchfactor\fontdimen3\font minus
  \fontdimen4\font\relax}
\providecommand{\BIBforeignlanguage}[2]{{%
\expandafter\ifx\csname l@#1\endcsname\relax
\typeout{** WARNING: IEEEtran.bst: No hyphenation pattern has been}%
\typeout{** loaded for the language `#1'. Using the pattern for}%
\typeout{** the default language instead.}%
\else
\language=\csname l@#1\endcsname
\fi
#2}}
\providecommand{\BIBdecl}{\relax}
\BIBdecl

\bibitem{SigtiaEtal:2018:IS}
\BIBentryALTinterwordspacing
S.~Sigtia, R.~Haynes, H.~Richards, E.~Marchi, and J.~Bridle, ``{Efficient Voice
  Trigger Detection for Low Resource Hardware},'' in \emph{Proc. Interspeech},
  Sept 2018, pp. 2092--2096. [Online]. Available:
  \url{http://dx.doi.org/10.21437/Interspeech.2018-2204}
\BIBentrySTDinterwordspacing

\bibitem{Wu:2018:ICASSP}
M.~{Wu}, S.~{Panchapagesan}, M.~{Sun}, J.~{Gu}, R.~{Thomas}, S.~N. {Prasad
  Vitaladevuni}, B.~{Hoffmeister}, and A.~{Mandal}, ``{Monophone-Based
  Background Modeling for Two-Stage On-Device Wake Word Detection},'' in
  \emph{2018 IEEE International Conference on Acoustics, Speech and Signal
  Processing (ICASSP)}, 2018, pp. 5494--5498.

\bibitem{MichaelyEtal:2017:ASRU}
A.~H. {Michaely}, X.~{Zhang}, G.~{Simko}, C.~{Parada}, and P.~{Aleksic},
  ``{Keyword Spotting for Google Assistant Using Contextual Speech
  Recognition},'' in \emph{2017 IEEE Automatic Speech Recognition and
  Understanding Workshop (ASRU)}, 2017, pp. 272--278.

\bibitem{MallidiEtal:2018:IS}
S.~Mallidi, R.~Maas, K.~Goehner, A.~Rastrow, S.~Matsoukas, and B.~Hoffmeister,
  ``{Device-directed Utterance Detection},'' in \emph{Proc. Interspeech}, Sept
  2018, pp. 1225--1228.

\bibitem{HuangEtal:2019:IS}
\BIBentryALTinterwordspacing
C.-W. Huang, R.~Maas, S.~H. Mallidi, and B.~Hoffmeister, ``{A Study for
  Improving Device-Directed Speech Detection Toward Frictionless Human-Machine
  Interaction},'' in \emph{Proc. Interspeech}, Sept 2019, pp. 3342--3346.
  [Online]. Available: \url{http://dx.doi.org/10.21437/Interspeech.2019-2840}
\BIBentrySTDinterwordspacing

\bibitem{LadhakEtal:2016:IS}
\BIBentryALTinterwordspacing
F.~Ladhak, A.~Gandhe, M.~Dreyer, L.~Mathias, A.~Rastrow, and B.~Hoffmeister,
  ``{LatticeRnn: Recurrent Neural Networks Over Lattices},'' in \emph{Proc.
  Interspeech}, Sept 2016, pp. 695--699. [Online]. Available:
  \url{http://dx.doi.org/10.21437/Interspeech.2016-1583}
\BIBentrySTDinterwordspacing

\bibitem{JeonLiuMason:2019:ICASSP}
W.~{Jeon}, L.~{Liu}, and H.~{Mason}, ``{Voice Trigger Detection from LVCSR
  Hypothesis Lattices Using Bidirectional Lattice Recurrent Neural Networks},''
  in \emph{2019 IEEE International Conference on Acoustics, Speech and Signal
  Processing (ICASSP)}, May 2019, pp. 6356--6360.

\bibitem{PranayEtal:2020:ICASSP}
P.~{Dighe}, S.~{Adya}, N.~{Li}, S.~{Vishnubhotla}, D.~{Naik}, A.~{Sagar},
  Y.~{Ma}, S.~{Pulman}, and J.~{Williams}, ``{Lattice-Based Improvements for
  Voice Triggering Using Graph Neural Networks},'' in \emph{ICASSP 2020 - 2020
  IEEE International Conference on Acoustics, Speech and Signal Processing
  (ICASSP)}, 2020, pp. {7459--7463}.

\bibitem{JabobsEtal:NeruComp:1991}
R.~Jacobs, M.~I. Jordan, S.~J. Nowlan, and G.~E. Hinton, ``{Adaptive Mixtures
  of Local Experts},'' \emph{Meual Computation}, February 1991.

\bibitem{Huangetal:ICASSP}
Z.~{Huang}, T.~{Ng}, L.~{Liu}, H.~{Mason}, X.~{Zhuang}, and D.~{Liu},
  ``{SNDCNN: Self-Normalizing Deep CNNs with Scaled Exponential Linear Units
  for Speech Recognition},'' in \emph{ICASSP 2020 - 2020 IEEE International
  Conference on Acoustics, Speech and Signal Processing (ICASSP)}, 2020, pp.
  6854--6858.

\bibitem{VeselyEtal:2013:IS}
K.~Vesel{\'y}, A.~Ghoshal, L.~Burget, and D.~Povey, ``{Sequence Discriminative
  Training of Deep Neural Networks},'' in \emph{Proc. Interspeech}, Aug 2013.

\bibitem{KlambauerEtal:2017:NIPS}
\BIBentryALTinterwordspacing
G.~Klambauer, T.~Unterthiner, A.~Mayr, and S.~Hochreiter, ``{Self-Normalizing
  Neural Networks},'' in \emph{Advances in Neural Information Processing
  Systems 30}, 2017, pp. 971--980. [Online]. Available:
  \url{http://papers.nips.cc/paper/6698-self-normalizing-neural-networks.pdf}
\BIBentrySTDinterwordspacing

\end{thebibliography}

\end{document}